\documentclass[preprint,prd,aps,showpacs]{revtex4}
\usepackage{graphicx}
\usepackage{latexsym}
\usepackage{epsfig}
\usepackage{axodraw4j}
\usepackage{color}
\usepackage{amsmath}
\begin{document}
\setcounter{page}{1}
\title{Fermion Self Energy Correction in Light-Front QED using  Coherent State Basis}
\author{Jai D. More\footnote{\tt more.physics@gmail.com}}
\author{Anuradha Misra\footnote{\tt misra@physics.mu.ac.in}}
\affiliation{Department of Physics,University of Mumbai,
\\Santa Cruz(E), Mumbai, India-400098}%

\begin{abstract}
We discuss Infrared(IR) divergences in lepton mass renormalization in Light-Front Quantum Electrodynamics (LFQED) in Feynman gauge. We consider LFQED with Pauli-Villars fields and using old-fashioned time ordered perturbation theory (TOPT),  we show that our earlier result regarding cancellation of true IR divergences up to $O(e^4)$ in coherent state basis holds in Feynman gauge also.
\end{abstract}
\date{\today}
\maketitle

\section{INTRODUCTION}
Light front Hamiltonian methods have been applied successfully to theories such as Yukawa theory, QED, $\phi^4$ theory and (1+1)-dimensional QCD to obtain the mass spectrum and wave functions \cite{ELLER87, KRA92,TANG91, HILLER93}. However, there are still unsolved issues which must be focused upon before one can develop non-perturbative methods for light-front quantum chromodynamics (LFQCD)\cite{WILSON94}. One of these important issues is the problem of infrared (IR) divergences \cite{WILSON94,HARI93}. In quantum electrodynamics, infrared divergences are eliminated if one uses appropriate initial and final states of charged particles with a suitable superposition of an infinite number of photons. The issue of cancellation of IR divergences at amplitude level has been addressed by Chung \cite{CHU65} and Kulish and Faddeev (KF)\cite{KUL70}. KF proposed a method of asymptotic dynamics wherein one replaces the free Hamiltonian by an asymptotic Hamiltonian that takes into account the long range interaction between incoming and outgoing states. This asymptotic Hamiltonian is used to construct an asymptotic evolution operator 
\begin{eqnarray}\label{omegadef}
\Omega^A_{\pm} = T~exp\bigg[-i\int_{\mp \infty}^0 V_{as}(t) dt \bigg]
\end{eqnarray}
and thereby a set of coherent states
\begin{equation} \label{cohstate}
\vert n;\pm \rangle = \Omega^A_{\pm}\vert n\rangle
\end{equation}
It was shown that the transition matrix elements formed using these asymptotic states are free of IR divergences.  Subsequently, KF method was discussed in the context of QCD by various authors \cite{BUT78, NEL81, NEL181, GRECO78, DAH81}. 

A coherent state formalism in light-front field theory (LFFT) has been discussed by various authors \cite{HARI88,VARY99}, in particular, in the context of cancellation of IR divergences in QED and QCD at lowest order \cite{ANU94,ANU96,ANU00}. One of us proposed the use of coherent state formalism in LFFT due to its usefulness in establishing the cancellation/ non-cancellation of IR divergences in Hamiltonian formalism \cite{ANU94, ANU96, ANU00, ANU05}. A coherent state formalism for LFQED in LF gauge was developed and applied to show the cancellation of IR divergence in one loop vertex correction \cite{ANU94}. Recently, we have extended the formalism  to demonstrate cancellation of IR divergences up to $O(e^4)$ in fermion self energy in Ref.~\cite{JAI12}, henceforth referred to as $\mathcal I$.

LF quantization is performed usually in light-front gauge $A^+=0$ due to its many advantages when applied to non-abelian theory \cite{PREM01,BRODSKY04}, in particular, the absence of ghost fields. One of the reasons for using LF gauge is that one is able to solve the constraint equation for the non-dynamical part of the fermion field. Gauge  independence of LFQED calculations has been addressed in literature. In particular, gauge dependence of non perturbative calculations has been studied and gauge invariance of physical quantities has been confirmed \cite{HILLER09}. In Ref. \cite{HILLER11}, mass eigenvalue problem in QED is discussed in an arbitrary covariant gauge. In this work, the theory is regulated by introducing  PV photons and electrons. It has been shown that in Feynman gauge, one PV electron and one PV photon are sufficient for canceling the instantaneous fermion interactions and the $A^+$ dependent terms cancel from the constraint equation\cite{HILLER11}.  

In $\mathcal I$, we have shown the cancellation of IR divergences up to $O(e^4)$ in LF gauge using coherent state basis. Here, we extend our analysis to show this cancellation in Feynman gauge. We calculate fermion self energy in LFQED up to $O(e^4)$ in Feynman gauge. We show that in Feynman gauge also the true IR divergences in lepton mass renormalization are cancelled up to $O(e^4)$ if one uses coherent state basis for evaluating the transition matrix elements. 
The plan of the paper is as follows: In Section II, we present the Hamiltonian of  PV regularized LFQED in general covariant gauge\cite{HILLER11} and calculate the $O (e^2)$ fermion mass renormalization using light-cone-time-ordered perturbation theory(LCTOPT) in Fock basis. We demonstrate the appearance of true IR divergences in the form of vanishing light-cone energy denominators. In Section III, we obtain the form of coherent states from this Hamiltonian using the method of asymptotic dynamics. In Section IV, we calculate $\delta m^2$ in lowest order using the coherent state basis and show that the extra contributions due to emission and absorption of soft photons indeed cancel the IR divergences in $\delta m^2$. In Section V, we calculate $\delta m^2$ up to $O(e^4)$ in Fock basis and identify the true IR divergences in it. In Section VI, we perform the same calculation in coherent state basis and show the cancellation of IR divergences. Section VII contains a summary and discussion of our results. Appendix A and B contain the details of the calculation of transition matrix elements in Fock basis and coherent state basis respectively.

\section{PRELIMINARIES}
\subsection{Light-Front QED in Feynman gauge}
We begin with the QED Lagrangian in Lorentz gauge with an
arbitrary gauge parameter $\zeta$ and additional PV fields \cite{HILLER11}:
\begin{eqnarray}  \label{Lagrangian}
{\cal L} &=&  \sum_{i=0}^2 (-1)^i \left[-\frac14 F_i^{\mu \nu} F_{i,\mu \nu} 
         +\frac{1}{2} \mu_i^2 A_i^\mu A_{i\mu} 
         -\frac{1}{2} \zeta \left(\partial^\mu A_{i\mu}\right)^2\right] \\
&& + \sum_{i=0}^2 (-1)^i \bar{\psi_i} (i \gamma^\mu \partial_\mu - m_i) \psi_i 
  - e \bar{\psi}\gamma^\mu \psi A_\mu . \nonumber
\end{eqnarray}
Here
\begin{equation} \label{NullFields}
  \psi =  \sum_{i=0}^2 \sqrt{\beta_i}\psi_i, \;\;
  A_\mu  = \sum_{i=0}^2 \sqrt{\xi_i}A_{i\mu}, \;\;
  F_{i\mu \nu} = \partial_\mu A_{i\nu}-\partial_\nu A_{i\mu} ,
\end{equation}
$i=0$ corresponds to the physical field and $i=1$ and 2, to PV fields. The fermion and photon fields have masses $m_i$ and $\mu_i$ respectively and $\mu_0=0$ for the physical photon. $\beta_i$ and $\xi_i$ are coupling coefficients which satisfy 
\begin{equation}
\sum_{i=0}^2(-1)^i\xi_i=0, \;\;
\sum_{i=0}^2(-1)^i\beta_i=0. \;\;
\end{equation}
The light-front Hamiltonian density, for a free massive vector field with mass $\mu $, is given by
\begin{align}\label{Hamdensity}
{\cal H}={\cal H}|_{\zeta=1}+\frac{1}{2}(1-\zeta)(\partial\cdot A)(\partial\cdot A -2\partial_-A_++2\partial_\perp A_\perp),
\end{align}
where the first term of the Eq.~(\ref{Hamdensity}) gives the Feynman gauge Hamiltonian density 
\begin{equation}
{\cal H}|_{\zeta=1}=\frac{1}{2}\sum_{\mu=0}^3 \epsilon^\mu
       \left[(\partial_\perp A^\mu)^2+\mu^2 (A^\mu)^2\right].
\end{equation}
Thus, the light-front QED Hamiltonian in Feynman gauge ($\zeta=1)$ is
\begin{equation}
P^-= H \equiv H_0 + V  \;,
\end{equation}
where the free Hamiltonian is given by
\begin{align} \label{QEDP-}
H_0=&\sum_{i=0}^2\sum_{s}\int d^2{\bf p}_\perp dp^+\frac{p_\perp^2+m_i^2}{2p^+}{(-1)}^i(b_{i}^\dagger (p,s) b_{i}(p,s)+d_{i}^\dagger (p,s) d_{i}(p,s)) \nonumber\\
     &+\sum_{l=0}^2\sum_{\lambda=0}^3\int [dk] \frac{k_\perp^2+\mu_{l\lambda}^2}{2k^+}{(-1)}^l \epsilon_l^\lambda a_{l}^\dagger(k,\lambda) a_{l}(k,\lambda)
\end{align}
and $V$ has the form 

\begin{align}\label{V}
V=e\int d^2 {\bf x}_\perp dx^-\int[dp][d\overline{p}][dk]\sum_{i,j,l}\sum_{s,s^\prime,\lambda} [e^{i\overline{p}\cdot x}\overline{u_i}(\overline{p},s^\prime)b_i^\dagger(\overline{p},s^\prime) +e^{-i\overline{p}\cdot x}\overline{v_i}(\overline{p},s^\prime)d_i(\overline{p},s^\prime)]\nonumber\\
\times\gamma^\mu[e^{-ip\cdot x} u_j(p,s) b_j(p,s)+e^{ip\cdot x} v_j(p,s)d_j^\dagger(p,s)] \epsilon^\lambda_{l\mu}(k)[e^{-ik\cdot x}a_l(k,\lambda)+e^{ik\cdot x}a_l^\dagger(k,\lambda)],
\end{align}
where
\begin{equation}
\int [dp] \equiv \int_{-\infty}^{\infty} {d^2{\bf p}_\perp \over {(2\pi)^{3\over 2}}} \int_0^\infty{dp^+ \over {\sqrt{2p^+}}}
\end{equation}
Summation over $i, j, l$ runs from $0$ to 2. Since PV-field contributions have been added to the Lagrangian, the instantaneous fermion terms cancel \cite{HILLER11}. The instantaneous photon terms that appear in light-cone gauge are also not present in Feynman gauge.  The polarization vectors $\epsilon_l^{(\lambda)}$ have an additional flavor index $l$, because they depend on the mass of the photon flavor. 

\subsection{Lepton Mass Renormalization in Light-Front QED}
In light-front time ordered perturbation theory, the transition matrix is given by the perturbative expansion
\begin{equation}
T= V + V {1 \over {p^--H_0}}V + \cdots
\end{equation}

The lepton mass shift is obtained by calculating $T_{pp}$ which is the matrix element of the above series between the initial and the final lepton states $\vert p,s \rangle$ and is given by
\begin{align}\label{deltam}
\delta m^2= p^+ \sum_{s} T_{pp}
\end{align}
We expand $T_{pp}$ in powers of $e^2$ as  
\begin{align}
T_{pp}=T^{(1)}+T^{(2)}+\cdots
\end{align}
In general, $T^{(n)}$ gives the $O(e^{2n})$ contribution to lepton self energy correction. Here, the initial (or final) lepton momentum is
\begin{align}
p=\biggl[p^+,\frac{{\bf p}_\perp^2+m^2}{2p^+},{\bf p}_\perp\biggr] ,
\end{align}
Momentum of internal photon line
\begin{align}
k=\biggl[k^+,\frac{{\bf k}_\perp^2+\mu_l^2}{2k^+},{\bf k}_\perp\biggr], \quad l=0, 1, 2
\end{align}
Momentum of the internal fermion line is 
\begin{align}
p_i=\biggl[p^+,\frac{{\bf p}_\perp^2+m_i^2}{2p^+},{\bf p}_\perp\biggr] , \quad i=0, 1, 2
\end{align}
$O(e^2)$ correction is obtained from
\begin{align}\label{def diag1}
T^{(1)}_{pp}\equiv T^{(1)}( p,p)=\langle p,s \vert V \frac{1} {p^- - H_0}V\vert p,s \rangle
\end{align}
Note that
\begin{align}
T^{(1)}_{pp} \equiv T^{(1)}(p,p)=T_{1a}+T_{1b}\;
\end{align}
where $T_{1a}$ and $T_{1b}$ are $O(e^2)$ contributions from standard three point vertices involving physical photon and PV photons  respectively and are represented by the diagrams in Fig. 1(a) and Fig. 1(b). 
\begin{figure}[h]
\includegraphics[scale=0.6]{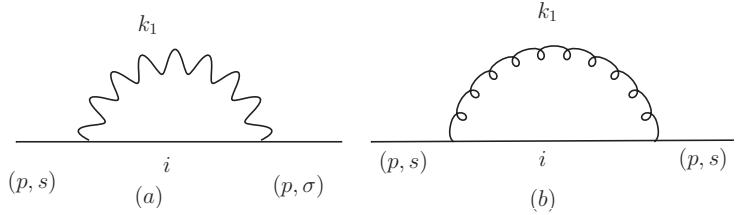}
\label{fig1}
\caption{Diagrams for $O(e^2)$ self energy correction in fock basis corresponding to $T_{1a}$ and $T_{1b}$. 
In Fig. (a), wavy line corresponds to physical photon $(i=0)$ and in Fig. (b) curly line corresponds to PV photon $(i=1,2)$.}
\end{figure}
The wavy arc represents physical photon while the spiral arc represents both PV photons $(j=1,2)$. The index i on the internal fermion line takes values $i=0$, 1 or 2, where $i=0$ corresponds to the  physical fermion field while $i=1$ and 2 represent PV fermion fields. We are interested in the true IR divergences which arise due to vanishing energy denominators in TOPT \cite{ANU94}.\\

The energy denominator in $T_{1b}$ is 
\begin{align}\label{denominator1}
p^--(p-k_1)_i^--k_{1j}^-=&\frac{p_\perp^2 +m^2}{2p^+}-\frac{(p_\perp-k_{1\perp})^2 +m_i^2+\mu_j^2}{2(p^+-k_1^+)}-\frac{k_{1\perp}^2 +\mu_j^2}{2k_1^+} 
=&-\frac{p\cdot k_1-m^2+ m_i^2}{p^+-k_1^+}
\end{align}
One can show that this energy difference is zero only when $m_i =m$. Therefore, up to $O(e^2)$ the energy denominator cannot be divergent if flavor changing vertices are involved. Moreover, due to non-zero mass of photon PV field, in the limit $k_1^+ \rightarrow 0$, ${\bf k}_{1\perp} \rightarrow 0$, $p\cdot k_1 \not\rightarrow 0$ irrespective of the flavor of fermion field. 
Thus $T_{1b}$ cannot have such IR divergences and we do not need this diagram for our discussion. Neglecting $T_{1b}$, $O(e^2)$ transition matrix element contributing to fermion self energy reduces to
\begin{eqnarray}
T_{1a}(p,p)=\langle p,s \vert V_1 \frac{1} {p^- - H_0}V_1\vert p,s \rangle
\end{eqnarray} 
Here $V_1$ is same as in $\mathcal I$  i.e. 3-point vertex involving physical fermion and physical photon $(i=0)$ only. Proceeding as in $\mathcal I$,
\begin{align}
\delta m^2_{1a}=\frac{e^2}{2(2 \pi)^3}\int{d^2{\bf k}_{1\perp}}\int\frac{dk_1^+}{k_1^+p_i^+} \frac{Tr[\epsilon\llap/^\lambda(k_1) (\not p_i+m) \epsilon\llap /^\lambda(k_1)(\not p+m)]}{4(p^--p_i^--k_1^-)}
\end{align}
where \quad $p_i=p-k_1$.\\
In the asymptotic limit, $k_1^+ \rightarrow 0$, ${\bf k}_{1\perp} \rightarrow 0$, we obtain 
\begin{eqnarray}\label{diag1}
{(\delta m^2_{1a})}^{IR}=-\frac{e^2}{(2\pi)^3}\int{d^2{\bf k}_{1\perp}}\int\frac{dk_1^+}{k_1^+}\frac{(p\cdot \epsilon(k_1))^2}{(p\cdot k_1)}
\end{eqnarray}
It is to be noted that the diagrams with PV photon fields and/ or flavor changing vertices do not contribute to IR divergences, as in both the cases the energy denominator is necessarily non-zero. Hence, we need to consider $i=0$ case only for Fig 1(a). The denominator vanishes in the limit $k_1^+ \rightarrow 0$, ${\bf k}_{1\perp} \rightarrow 0$ i.e $p\cdot k_1 \rightarrow 0$, thus leading to IR divergences \cite{ANU94,JAI12}. 

It was pointed out in Ref.~\cite{HILLER11} that in the infinite PV mass limit a tree level diagram involving an intermediate PV fermion of mass $m_1$ reduces to four point instantaneous vertex  as illustrated in Fig. 2. 
\begin{figure}[h]
\includegraphics[scale=0.7]{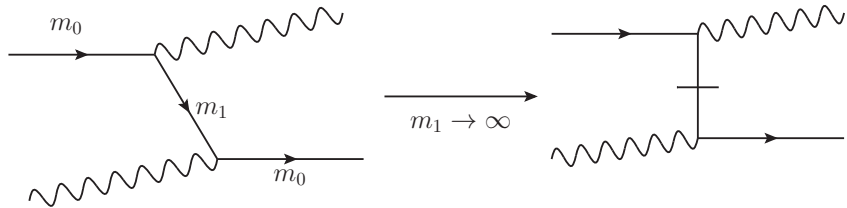}
\label{PV1}
\caption{In the infinite PV mass limit the PV fermion line reduces to an instantaneous four point  interaction term denoted by a dash on the fermion line .}
\end{figure}
Similarly, Fig.1(a) with $i = 1$, reduces to the instantaneous diagram in  Fig. 1(b) of $\mathcal I$ in the infinite  PV fermion mass limit. However, such diagrams 
do not  contain a vanishing energy denominator and hence are not relevant for the present discussion.

\section{COHERENT STATE FORMALISM AND INFRARED DIVERGENCES}
In $\mathcal I$, it was shown that the true IR divergences in self energy correction up to $O(e^4)$ get cancelled if one uses coherent state basis in LFQED in LF gauge. We will prove the same result in Feynman gauge here. For that purpose, we will now obtain the form of coherent states for LFQED in Feynman gauge by the method used in Ref.\cite{KUL70} for equal time theory. As shown in Ref.\cite{HILLER11}, when one writes the Hamiltonian in terms of independent degrees of freedom, the non-local terms do not appear and hence, in this formalism, there are no instantaneous interaction terms in LF Hamiltonian.\\
\indent The light-front time dependence of the interaction Hamiltonian is given by \cite{ANU94}
\begin{align}\label{interactionH}
V(x^+)= e\sum_{\alpha=1}^4 \int d\nu_\alpha[ e^{-i \nu_\alpha x^+} {\tilde h}_\alpha(\nu_\alpha)
+ e^{i \nu_\alpha x^+} {\tilde h}^{\dagger}_\alpha (\nu_\alpha)]
\end{align}
 ${\tilde h}_\alpha(\nu_\alpha)$ are three point interaction vertices. For example, 
\begin{equation}
{\tilde h}_1 = \sum_{i,j,l=0}^2\sum_{s\lambda} b_i^{\dagger}(\overline p,s^\prime)b_j(p,s) a_l(k,\lambda)  \overline{u_i}(\overline{p},s^\prime)\gamma^\mu u_j(p,s)\epsilon_{l\mu}^{(\lambda)}(k);,
\end{equation}
and
$\nu_\alpha$ is  the light-front energy transferred at the vertex ${\tilde h}_\alpha$.  For example,
\begin{equation}\label{nu1}
\nu_1 = p_j^- + k_l^- - \overline p_i^- = {p_j \cdot k_l \over p^++k^+}
\end{equation}
is the energy transfer at $ee\gamma$ vertex. The integration measure is given by
\begin{equation}
\int d\nu={1\over{{(2\pi)}^{3/2}}}\int{{[dp][dk]}\over{\sqrt{2\overline p^+}}} \;
\end{equation}
$\overline p^+$ and $\overline {\bf p}_{\perp}$ being fixed at each vertex by momentum
conservation.\\
\indent In the limits $\left|x^+\right|\rightarrow \infty$, non-zero contributions to $V(x^+)$ come only from regions where $\nu_\alpha \rightarrow 0$. It is easy to see that $\nu_2$ and $\nu_3$ are always non-zero and therefore, $\tilde h_2$ and $\tilde h_3$ do not appear in the asymptotic Hamiltonian. As	 shown earlier in Eq.(\ref{denominator1}) the energy differences  $\nu_1$ and $\nu_4$ cannot be zero at flavor changing vertices or at vertices involving massive photon fields. This means $\nu_1$ and $\nu_4$ can be zero only for the physical photons and fermions but not for PV photons and fermions. Thus, the IR divergent contribution to asymptotic Hamiltonian comes only from the terms with $i=j=l=0$ in Eq.~(\ref{interactionH}). Thus, the 3-point asymptotic Hamiltonian is defined by the following expression \cite {ANU94}
\begin{eqnarray}\label{Vas}
V_{as}(x^+)=e\sum_{\alpha=1,4}\int d\nu_\alpha \Theta_\Delta(k)[e^{-i \nu_\alpha x^+}\tilde h_\alpha^{(0)}(\nu_\alpha) +e^{i \nu_\alpha x^+}\tilde h^{(0)\dagger}_\alpha(\nu_\alpha)] \;
\end{eqnarray} 
where $\Theta_{\Delta}(k)$ is a function which takes value 1 in the asymptotic region and is zero elsewhere. The superscript on $\tilde h_\alpha^{(0)}(\nu_\alpha)$ indicates the $i=j=l=0$ part of $V(x^+)$. Eq.~(\ref{Vas}) is same as $V_{as}(x^+)$ in Ref. \cite{ANU94} and 3-point vertex part of  $V_{as}(x^+)$ in $\mathcal I$. 
The detailed  procedure for obtaining $V_{as}(x^+)$ can be found in the Refs.\cite{ANU94,JAI12}. The asymptotic states are obtained from the asymptotic Hamiltonian:
\begin{equation} 
\vert n;\pm \rangle = \Omega^A_{\pm}\vert n\rangle
\end{equation}
where $\Omega^A_{\pm} $ is the asymptotic evolution operator and $\vert n\rangle$ is the Fock state. $\Omega^A_{\pm} $ is defined by 
\begin{eqnarray}
\Omega^A_{\pm} = T~exp\bigg[-i\int_{\mp \infty}^0 V_{as}(x^+) dx^+ \bigg]
\end{eqnarray}
Following the procedure in $\mathcal I$, the asymptotic states are found to be 
\begin{align}\label{omega}
\Omega_{\pm}^A \vert n \rangle=&exp\biggl[-e\int{dp^+d^2{\bf p}_\perp}\int\sum_{\lambda=1,2}
\frac{d^2{\bf k}_\perp}{(2\pi)^{3/2}}\int{{dk^+}\over{\sqrt{2k^+}}}\nonumber\\ &[f(k,\lambda:p)a^\dagger(k,\lambda)-f^*(k,\lambda:p)a(k,\lambda)]\rho(p)\biggr]\vert n  \rangle \;
\end{align}
Here
\begin{equation}
\rho(p) = \sum_{i=0}^2(b_i^\dagger(p) b_i(p)-d_i^\dagger(p) d_i(p)), 
\label{rho}
\end{equation}
\begin{equation}
a(k, \lambda) \equiv a_{0}(k, \lambda)
\end{equation}
and
\begin{equation}
f(k,\lambda \colon p) = {{p_\mu\epsilon_\lambda^\mu(k)} \over {p\cdot k}}
\theta\bigg(\frac{k^+\Delta}{p^+}-{\bf k}_\perp^2\bigg)
\theta\bigg(\frac{p^+\Delta}{m^2}-k^+\bigg) \;,
\label{softphoton}
\end{equation}
One must notice that the form of asymptotic state is simpler here as compared to that in LF gauge (discussed in  $\mathcal I$) as 4-point instantaneous interaction does not appear in Feynman gauge Hamiltonian. Furthermore, there is no contribution to asymptotic Hamiltonian from PV fields as they are massive and hence $p^--k^--(p-k)^-\not\rightarrow 0$ in this case. One can use Eqs.~(\ref{cohstate}), (\ref{omega}) and (\ref{rho}) to obtain the form of coherent states
\begin{align}
\Omega_{\pm}^A \vert n  \rangle=
&exp\biggl[-e\int \sum_{\lambda=1,2}
\frac{d^2{\bf k}_\perp}{(2\pi)^{3/2}}\int \frac{dk^+}{\sqrt{2k^+}}[f(k,\lambda,p) a^\dagger(k,\lambda) - f^*(k,\lambda,p)a(k,\lambda)]\biggr]\vert n  \rangle \;
\end{align}
\vskip 0.5cm
\section{LEPTON MASS RENORMALIZATION UP TO $O(e^2)$ IN COHERENT STATE BASIS}
\begin{figure}[h]
\includegraphics[scale=0.6]{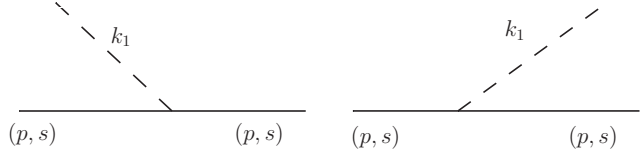}
\label{fig3}
\caption{Additional diagrams in coherent state basis for $O(e^2)$ self energy correction corresponding to $T^\prime$}
\end{figure}
The self energy contribution up to $O(e^2)$ in coherent state basis is given by $T^{(1)}+T^{^\prime(1)}$, where $T^{(1)}$ is defined in Eq.(\ref{def diag1}) and $T^{\prime(1)}$ arises from $O(e^2)$ term in 
\begin{align}\label{coh}
T^\prime(p,p)=&\langle p,s \colon f(p) \vert V\vert p,s \colon f(p) \rangle
\end{align}
and is represented by Fig. 3. A soft photon in coherent state is shown by the dotted line in the Feynman diagram in Fig. 3. The detailed calculation has been discussed in $\mathcal I$. Here we present only the result:
\begin{align}\label{diag2}
{(\delta m^2)}^\prime=\frac{e^2}{(2 \pi)^3} \int{d^2{\bf k}_{1\perp}}\int\frac{dk_1^+}{k_1^+}\frac{(p\cdot \epsilon(k_1))^2 \Theta_\Delta (k_1)}{p\cdot k_1}
\end{align}
where the prime indicates the correction due to additional terms in coherent state basis. \\
The energy denominator in Eq.~(\ref{diag2}) vanishes in the limit $k_1^+ \rightarrow 0,{\bf k}_{1\perp}\rightarrow 0$ thus leading to IR divergences. Adding  Eqs.~(\ref{diag1}) and (\ref{diag2}), these true IR divergences get cancelled and the $O(e^2)$ lepton mass correction is IR divergence free.
\vskip 0.5cm
\section{LEPTON MASS RENORMALIZATION UPTO $O(e^4)$ IN FOCK BASIS}
We will now calculate $O(e^4)$ lepton mass correction in Fock basis. Transition matrix element for $O(e^4)$ correction to self energy is given by
\begin{eqnarray}
T^{(2)}=T_{4}                 
\end{eqnarray}
where
\begin{align}\label{T_3}
T_{4}=&\langle p,s \vert V \frac{1} {p^- - H_0}V \frac{1} {p^- - H_0}V \frac{1} {p^- - H_0}V \vert p,s \rangle \\ \label{T_4}
\end{align}
Note that due to presence of PV fields, we get an additional set of diagrams involving 3-point vertices in addition to diagrams in $\mathcal I$. All these diagrams can be evaluated in the standard manner by inserting appropriate number of complete sets of intermediate states. Moreover, in present formalism, the four point instantaneous interaction term are absent in the Hamiltonian and therefore diagrams in Figs. 4, 5 and 6 in $\mathcal I$ are not present here. We will not reevaluate the diagrams already evaluated in $\mathcal I$  and discuss here only the additional diagrams which appear due to introduction of PV fields. We give the details of the calculation in Appendix A and present here only the results for the new diagrams.
\begin{figure}[h]
\includegraphics[scale=0.6]{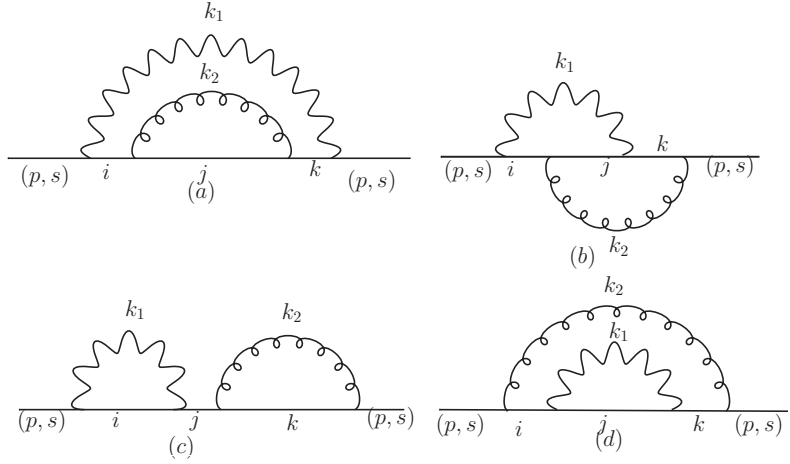}
\label{fig4}
\caption{Additional diagrams for $O(e^4)$ self energy correction in fock basis corresponding to $T_4$ (Complete set of diagrams consists of these and Figure 3 in $\mathcal I$)}
\end{figure}
Additional contribution to $T_4$ due to PV fields is given by
\begin{align}
T_{4}^{PV}\equiv T_4^{PV}(p,p)=T_{4a}^{PV}+T_{4b}^{PV}+T_{4c}^{PV}+T_{4d}^{PV}
\end{align}
where $T_{4a}^{PV}$, $T_{4b}^{PV}$, $T_{4c}^{PV}$ and $T_{4d}^{PV}$ correspond to Figs. 4(a)-(d) and are given by Eqs.~(\ref{T_3a}), (\ref{T_3b}) and (\ref{T_3c}). Using expressions for energy denominator from Eqs.~(\ref{relation1}) and (\ref{relation2}) we obtain
\begin{align}
(\delta m^2)_{4a}^{PV}=&-\frac{e^4}{2(2\pi)^6} \int {{d^2 {\bf k}_{1\perp}}{d^2{\bf k}_{2\perp}}}\int {{{dk_1^+}\over{k_1^+}}{{dk_2^+}\over {k_2^+p_i^+p_j^+p_k^+}}}\nonumber\\ &\frac{Tr[\not\epsilon^{\lambda_1} (k_1)(\not p_k+m_k)\not\epsilon^{\lambda_2}(k_2)(\not p_j+m_j) \not\epsilon^{\lambda_2} (k_2)(\not p_i+m_i) \not\epsilon^{\lambda_1}(k_1)(\not p+m)]} {32(p^--p_k^--k_1^-)(p^--p_i^--k_1^-)(p^--p_j^--k_1^--k_2^-)}
\end{align}
\begin{align}
(\delta m^2)_{4b}^{PV}=&-\frac{e^4}{2(2\pi)^6} \int {{d^2 {\bf k}_{1\perp}}{d^2{\bf k}_{2\perp}} }\int {{{dk_1^+}\over{k_1^+}}{{dk_2^+}\over {k_2^+p_i^+p_j^+p_k^+}}}\nonumber\\ &\frac{Tr[\not\epsilon^{\lambda_2} (k_2)(\not p_k+m_k)\not\epsilon^{\lambda_1}(k_1)(\not p_j+m_j)\not\epsilon^{\lambda_2}(k_2)(\not p_i+m_i) \not\epsilon^{\lambda_1}(k_1)(\not p+m)]}{32(p^--p_k^--k_2^-)(p^--p_i^--k_1^-)(p^--p_j^--k_1^--k_2^-)}
\end{align}
\begin{align}
(\delta m^2)_{4c}^{PV}=&-\frac{e^4}{2(2\pi)^6} \int {{d^2 {\bf k}_{1\perp}}{d^2{\bf k}_{2\perp}} }\int {{{dk_1^+}\over{k_1^+}}{{dk_2^+}\over {k_2^+p_i^+p_j^+p_k^+}}}\nonumber\\ &\frac{Tr[\not\epsilon^{\lambda_2} (k_2)(\not p_k+m_k)\not\epsilon^{\lambda_2}(k_2)(\not p^{\prime}_j+m_j) \not\epsilon^{\lambda_1}(k_1) (\not p_i+m_i)\not\epsilon^{\lambda_1}(k_1)(\not p+m)]} {32(p^--p_k^--k_2^-)(p^--p_i^--k_1^-)(p^--p^{\prime-}_j)}
\end{align}
where $p_j^\prime =p$ and $p_i$, $p_j$ and $p_k$ have been defined in Eqs.~(\ref{pi}), (\ref{pj}) and (\ref{pk}) respectively.
The last term $T_{4d}^{PV}$ is IR convergent, as the energy denominator involved can never be zero, and hence it is not needed for our discussion.
Note that the divergence structure of these diagrams and those in $\mathcal I$ is different i.e 
$(\delta m^2)_{4a}^{PV}$, $(\delta m^2)_{4b}^{PV}$ and $(\delta m^2)_{4c}^{PV}$ can have IR divergences only when $p\cdot k_1 \rightarrow 0$ i.e $k_1^+\rightarrow 0, {\bf k}_{1\perp}\rightarrow 0$, $i=j=k=0$ and $l=0$ i.e. only when the diagram involves physical fermions and photons. Also, as the PV field is massive, the energy denominator involving its $k^-$ momentum cannot be zero and therefore we need not consider the limit $k_2^+\rightarrow 0, {\bf k}_{2\perp}\rightarrow 0$ and the combined limit $k_1^+\rightarrow 0, {\bf k}_{1\perp}\rightarrow 0, k_2^+\rightarrow 0, {\bf k}_{2\perp}\rightarrow 0$ as we did in $\mathcal I$.
As shown in Appendix A, the IR contribution from diagrams in Figs. 4(a) and 4(b) is given by
\begin{align}\label{3a}
&(\delta m^2)_{4a}^{PV}+(\delta m^2)_{4b}^{PV}\nonumber\\
&=-{e^4\over{(2\pi)^6}}\int{{d^2{\bf k}_{1\perp}}{d^2{\bf k}_{2\perp}}}\int{{{dk_1^+}\over{k_1^+}}{{dk_2^+}\over {k_2^+}}}\frac{[2(p\cdot\epsilon(k_1))^2(p\cdot \epsilon(k_2))^2-(p\cdot k_2)(p\cdot \epsilon(k_1))^2]}{4(p\cdot k_1)^2(p\cdot k_2)}
\end{align}
and the contribution from Fig. 4(c) is given by
\begin{align}\label{3c}
(\delta m^2)_{4c}^{PV}=&{e^4\over{(2 \pi)^6}} \int {{d^2{\bf k}_{1\perp}}{d^2{\bf k}_{2\perp}} }\int {{{dk_1^+}\over{k_1^+}}{{dk_2^+}\over {k_2^+}}}\frac{2(p\cdot \epsilon(k_1))^2(p\cdot \epsilon(k_2))^2-(p\cdot k_2)(p\cdot \epsilon(k_1))^2}{8p^+}\nonumber\\&\biggl[\frac{p^+}{(p\cdot k_1)^2(p\cdot k_2)}+\frac{p_k^+}{(p\cdot k_1)(p\cdot k_2)^2}\biggr]
\end{align}
Here we have used Heitler method \cite{HEIT54} discussed in Appendix A to deal with the vanishing denominator $(p^--p^{\prime-})$.

In addition to diagrams in Fig.4, there are $O(e^4)$ diagrams containing only physical photons but PV fermions in the intermediate states. These will appear the same as diagrams in  Fig. 3(a) of  $\mathcal I$ except that the internal fermion lines will correspond to PV fermions. One can draw these diagrams by replacing the curly lines by wavy lines in Fig. 4(a)-(c) here. Such diagrams will not involve vanishing energy denominators due to the presence of flavor changing interactions as discussed in Section II. However, in the infinite  PV fermion mass limit, such diagrams will reduce to diagrams involving  instantaneous interaction vertex. This is  illustrated schematically in Fig. 5, where Fig. 3(a) of $\mathcal I$ with the intermediate term involving PV fermion field reduces to Fig. 4(a) of $\mathcal I$ in infinite PV fermion mass limit. 

\begin{figure}[h]
\includegraphics[scale=1]{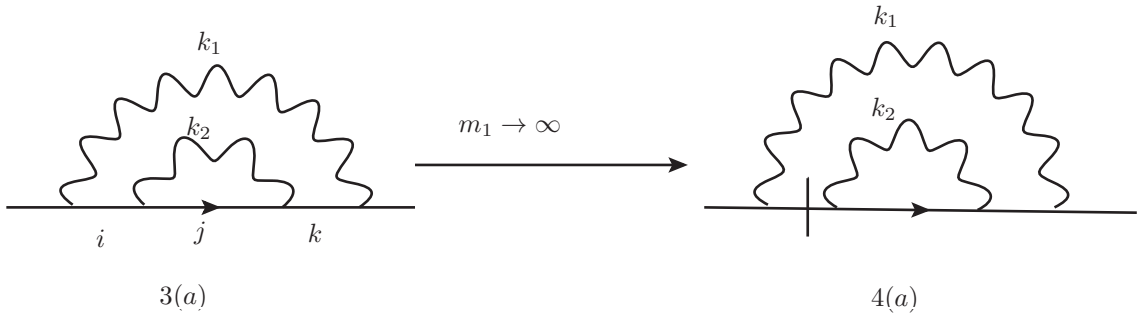}
\label{PV3}
\caption{In the infinite PV mass limit the diagram on the left reduces to a diagram involving instantaneous interaction. Here $i = 1$ while $j=k=0$. This figure involves intermediate PV fermion  also while Fig. 3(a) of $\mathcal I$ involved only the physical fermion terms.}
\end{figure}
\vskip 0.25cm 
\section{LEPTON MASS RENORMALIZATION IN COHERENT STATE BASIS UP TO $O(e^4)$}
In this section, we calculate the $O(e^4)$ lepton mass correction using coherent state basis. We will show that the IR divergent contribution from the additional diagrams in  coherent state basis exactly cancel the IR divergences arising due to vanishing energy denominators calculated in Section V. In coherent state basis, $O(e^4)$ correction to self energy is given by
\begin{align}
T^{(2)}+T_4^\prime \nonumber
\end{align}
where $T_4^\prime$ is $O(e^4)$ term in $\langle p,s \colon f(p) \vert V \frac{1} {p^- - H_0}V\frac{1} {p^- - H_0}V\vert p,s \colon f(p) \rangle$ represented by Fig. 6 here and Fig. 4 in $\mathcal I$. We present the details of calculation in Appendix B and give below only the contribution of Fig. 6 to $(\delta m^2)^\prime$. Cancellation of divergences in diagrams involving only massless photon has been given in $\mathcal I$ and we will not repeat it here.
\begin{figure}[h]
\includegraphics[scale=0.6]{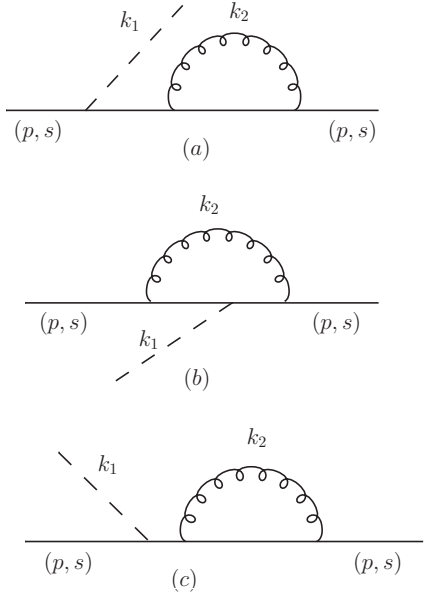}
\label{fig6}
\caption{Additional diagrams in coherent state basis for $O(e^4)$ self energy correction corresponding to $T_4^\prime$}
\end{figure}
The contribution of Fig. 6 is given by
\begin{equation}
{(\delta m^2)}_{4}^\prime={(\delta m^2)}_{6a}^\prime+{(\delta m^2)}_{6b}^\prime+{(\delta m^2)}_{6c}^\prime
\end{equation}
$(\delta m^2)_{6a}^\prime$, $(\delta m^2)_{6b}^\prime$ and $(\delta m^2)_{6c}^\prime$ have been evaluated in Appendix B and we give below only the result, 
\begin{align}\label{coh4a}
{(\delta m^2)}_{6a}^\prime=&\frac{e^4}{(2\pi)^6}\int{{d^2{\bf k}_{1\perp}}{d^2{\bf k}_{2\perp}}}\int{{{dk_1^+} \over{k_1^+}}{{dk_2^+}\over{k_2^+}}}\nonumber\\ &\frac{[2(p\cdot\epsilon(k_1))^2(p\cdot\epsilon(k_2))^2-(p\cdot k_2)(p\cdot \epsilon(k_1))^2]}{4(p\cdot k_1)^2[(p\cdot k_1)+(p\cdot k_2)-(k_1\cdot k_2)]} \Theta_{\Delta}(k_1)
\end{align}
\begin{align}\label{coh4b}
{(\delta m^2)}_{6b}^\prime=&{e^4\over{(2 \pi)^6}}\int{{d^2{\bf k}_{1\perp}}{d^2{\bf k}_{2\perp}}}\int{{{dk_1^+} \over{k_1^+}}{{dk_2^+}\over{k_2^+}}}\nonumber\\ &\frac{[2(p\cdot\epsilon(k_1))^2(p\cdot\epsilon(k_2))^2]}{4(p\cdot k_1)(p\cdot k_2)[(p\cdot k_1)+(p\cdot k_2)-(k_1\cdot k_2)]}\Theta_{\Delta}(k_1)
\end{align}
Adding Eqs.~(\ref{coh4a}) and (\ref{coh4b}) we obtain
\begin{align}\label{coh4}
(\delta m^2)_{6a}^\prime+(\delta m^2)_{6b}^\prime=&{e^4\over{(2 \pi)^6}}\int{{d^2{\bf k}_{1\perp}}{d^2{\bf k}
_{2\perp}}}\int{{{dk_1^+}\over{k_1^+}}{{dk_2^+}\over{k_2^+}}}\nonumber\\
&\frac{[2(p\cdot\epsilon(k_1))^2(p\cdot \epsilon(k_2))^2-(p\cdot k_2)(p\cdot \epsilon(k_1))^2]}{4(p\cdot k_1)^2(p\cdot k_2)} \Theta_{\Delta}(k_1)
\end{align}
Also
\begin{align}\label{coh4c}
(\delta m^2)_{6c}^\prime=&-{e^4\over{(2\pi)^6}}\int{{d^2{\bf k}_{1\perp}}{d^2{\bf k}_{2\perp}}}\int{{{dk_1^+} \over{k_1^+}}{{dk_2^+}\over{k_2^+}}}\frac{[2(p\cdot\epsilon(k_1))^2(p\cdot \epsilon(k_2))^2-(p\cdot k_2)(p\cdot \epsilon(k_1))^2]}{8p^+}\nonumber\\&\biggl[\frac{p^+}{(p\cdot k_1)^2(p\cdot k_2)}+\frac{p_3^+}{(p\cdot k_2)^2(p\cdot k_1)}\biggr]\Theta_{\Delta}(k_1)
\end{align} 
Adding Eqs.~(\ref{3a}) and (\ref{coh4}) we see that $(\delta m^2)_{4a}^{PV}+(\delta m^2)_{4b}^{PV}+(\delta m^2)_{6a}^\prime+(\delta m^2)_{6b}^\prime$ is IR finite.\\
Similarly, adding Eqs.~(\ref{3c}) and (\ref{coh4c}) we find $(\delta m^2)_{4c}^{PV} +(\delta m^2)_{6c}^\prime$ is IR finite. Adding all the contributions coming from Figs. 4 and 6 and combining with the results in $\mathcal I$, we find that the self energy correction up to $O(e^4)$ is IR finite. This completes the proof of cancellation of true IR divergences up to $O(e^4)$ for fermion self energy correction in coherent state basis in Feynman gauge also. 
\vskip 0.5cm
\section{CONCLUSION}
\vskip 0.5cm
 In $\mathcal I$, we have calculated lepton self energy correction in light-front QED in LC gauge up to $O(e^4)$ and have shown that the {\it true} IR divergences get cancelled when coherent state basis is used to calculate the matrix elements.  In this work, we have obtained the same result in Feynman gauge. One can notice that the proof of cancellation is simpler in Feynman gauge. The proof can be generalized to general covariant gauges. We plan to address this in a future work. 

The cancellation of IR divergences between real and virtual processes is known to hold in equal-time QED to all orders. This cancellation was also shown by Kulish and Faddeev \cite{KUL70} using the coherent state formalism. KF method leads to cancellation of IR divergences in QED to all orders. However, it is well known that the Bloch-Nordseick theorem \cite{BLOCH37} does not hold in QCD and therefore, in QCD one does not expect to construct an all order proof of cancellation of IR divergences along the lines of KF method. Basically, the non-cancellation of IR divergences in QCD arises due to the fact that asymptotic states here are bound states of quarks and anti-quarks and therefore the asymptotic Hamiltonian to be used in KF method should contain the confining potential and should not be just the asymptotic Hamiltonian of QCD. An "improved" method of asymptotic dynamics has been introduced by McMullan {\it etal} \cite{HOR98, HOR99, HOR00} which takes into account the separation of particles also. The improved method has also been discussed in the context of LFQED and LFQCD\cite{ANU05}.
 
An improved coherent state method in LFFT may be useful from the point of view of extracting information about the artificial confining potential which is needed in LF bound state calculations\cite{WILSON94}. If we use appropriate Hamiltonian of bound states as the asymptotic Hamiltonian and develop a coherent state approach based on it, then this approach would lead to cancellation of IR divergences in QCD as well. It will be interesting to understand this connection between cancellation/non-cancellation of IR divergences and the form of asymptotic Hamiltonian. Thus we may be able to get some perception of the form of artificial confining potential mentioned by Wilson et. al. in Ref. \cite{WILSON94} by understanding the structure of IR divergences in LFQCD.
\vskip 0.25cm
{\bf ACKNOWLEGEMENTS}
\vskip 0.25cm
A. Misra would like to thank BRNS under the Grant No. 2010/37P/47/BRNS and Jai More would like to thank Department of Science and Technology, India under the Grant No. SR/S2/HEP-17/2006 for the financial support.

\appendix 
\vskip .25cm
\section{Transition matrix element for self energy in fock basis}
We will need the following expressions for energy denominators:
\begin{align}\label{relation1}
p^--k_1^--(p-k_1)^-=&-\frac{(p\cdot k_1)}{p^+-k_1^+}\\
\label{relation2}
p^- - k_1^-- k_2^- -(p-k_1-k_2)^- =&-\frac {p \cdot k_1+p \cdot k_2-k_1 \cdot k_2}{p^+-k_1^+-k_2^+}
\end{align}

We now calculate $T_4$ which is defined by Eq.~(\ref{T_3}) and corresponds to  Fig. 4. Inserting complete sets of intermediate states in $T_{4}$, we obtain
\begin{align}
T_4(p,p) =T_{4a}^{PV}+T_{4b}^{PV}+T_{4c}^{PV}+T_{4d}^{PV}\nonumber
\end{align}where
\begin{align}\label{T_3a}
T_{4a}^{PV}=&{e^4\over{(2 \pi)^6}}\int{{d^2 {\bf k}_{1\perp}}{d^2{\bf k}_{2\perp}}\over 2p^+}\int{{dk_1^+}{dk_2^+}\over{32 k_1^+k_2^+p_i^+p_j^+p_k^+}}\nonumber\\&
\frac{\overline u(p,s)[\not\epsilon^{\lambda_1}(k_1)(\not p_k+m_k) \not\epsilon^{\lambda_2}(k_2)(\not p_j+m_j)\not\epsilon^{\lambda_2}(k_2)(\not p_i+m_i)\not\epsilon^{\lambda_1}(k_1)]u(p,s)} {(p^--p_k^--k_1^-)(p^--p_i^--k_1^-)(p^--p_j^--k_1^--k_2^-)}
\end{align}
with
\begin{align}\label{pi}
p_i=p-k_1,\\
\label{pj}
p_j=p-k_1-k_2.
\end{align}
Note that, here $p_i=p_k$\\
Similarly,
\begin{align}\label{T_3b}													
T_{4b}^{PV}=&{e^4\over{(2 \pi)^6}}\int {{d^2 {\bf k}_{1\perp}}{d^2 {\bf k}_{2\perp}}\over 2p^+ }\int{{dk_1^+}{dk_2^+}\over{32 k_1^+k_2^+p_i^+p_j^+p_k^+}}\nonumber\\&\frac{\overline u(p,s)[\not\epsilon^{\lambda_2}(k_2)(\not p_k+m_k)\not\epsilon^{\lambda_1}(k_1)(\not p_j+m_j) \not\epsilon^{\lambda_2}(k_2)(\not p_i+m_i)\not\epsilon^{\lambda_1}(k_1)] u(p,s)} {(p^--p_k^--k_2^-)(p^--p_i^--k_1^-)(p^--p_j^--k_1^--k_2^-)}
\end{align}
with
\begin{align}\label{pk}
p_k=p-k_2
\end{align}
\begin{align}\label{T_3c}
T_{4c}^{PV}=&\frac{e^4}{(2\pi)^6}\int{{d^2 {\bf k}_{1\perp}}{d^2 {\bf k}_{2\perp}}\over 2p^+ } \int {{dk_1^+}{dk_2^+}\over {32 k_1^+k_2^+p_i^+p_j^+p_k^+}}\nonumber\\&\frac{\overline u(p,s)[\not\epsilon^{\lambda_2}(k_2)(\not p_k+m_k) \not\epsilon^{\lambda_2}(k_2) (\not p_j^\prime+m) \not\epsilon^{\lambda_1}(k_1)(\not p_i+m_i)\not\epsilon^{\lambda_1}(k_1)] u(p,s)}{(p^--p_i^--k_1^-)(p^--p_j^{\prime -})(p^--p_k^--k_2^-)}
\end{align}
In the limit, $k_1^+\rightarrow 0, {\bf k}_{1\perp}\rightarrow 0$ and $i=j=k=0$, we get the IR divergent contribution. In this limit, Eqs.~(\ref{T_3a}) and (\ref{T_3b}) can be added such that the denominator reduces to $(p\cdot k_1)^2(p\cdot k_2)$ 
\begin{align}\label{delta66}
(\delta m^2)_{4a}^{PV}+(\delta m^2)_{4b}^{PV}=&\frac{e^4}{2(2\pi)^6}\int{{d^2 {\bf k}_{1\perp}}{d^2{\bf k}_{2\perp}}}\int {{dk_1^+}{dk_2^+}\over{32 k_1^+k_2^+}}\nonumber\\&\frac{Tr[\not\epsilon^{\lambda_1}(k_1)(\not p+m) \not\epsilon^{\lambda_2}(k_2)(\not p_k+m)\not\epsilon^{\lambda_2}(k_2)(\not p+m)\not\epsilon^{\lambda_1}(k_1)(\not p+m)]}{(p\cdot k_1)^2(p\cdot k_2)}
\end{align}
After calculating the trace, Eq.~(\ref{delta66}) leads to
\begin{align}
& (\delta m^2)_{4a}^{PV}+(\delta m^2)_{4b}^{PV}\nonumber\\
&={e^4\over{(2\pi)^6}}\int{{d^2{\bf k}_{1\perp}}{d^2{\bf k}_{2\perp}}}\int{{{dk_1^+}\over{k_1^+}}{{dk_2^+}\over {k_2^+}}}\frac{[2(p\cdot \epsilon(k_1))^2(p\cdot \epsilon(k_2))^2-(p\cdot k_2)(p\cdot\epsilon(k_1))^2]}{4(p\cdot k_1)^2(p\cdot k_2)}
\end{align}
Similarly, $T_{4c}^{PV}$ leads to 
\begin{align}\label{delta3c}
\delta m^2_{4c}=&\frac{e^4}{2(2\pi)^6}\int{{d^2 {\bf k}_{1\perp}}{d^2 {\bf k}_{2\perp}}}\int {{dk_1^+}{dk_2^+}\over {32k_1^+k_2^+p_i^+p_j^+p_k^+}}\nonumber\\ &\frac{Tr[\not\epsilon^{\lambda_2}(k_2)(\not p_k+m) \not\epsilon^{\lambda_2}(k_2) (\not p_j^\prime+m)\not\epsilon^{\lambda_1}(k_1)(\not p_i+m)\not\epsilon^{\lambda_1}(k_1)(\not p+m)]} {(p^--p_i^--k_1^-)(p^--p_2^{\prime -})(p^--p_k^--k_2^-)}
\end{align}
where $p_i$ and $p_k$ are defined by Eqs.~(\ref{pi}) and (\ref{pk}) and $p_j^\prime=p$. Note that  this  diagram is one-particle reducible, and therefore the energy denominator associated with the single-particle state vanishes. We have used the Heitler method \cite{HEIT54} for evaluating all such integrals. Using this method, we write \cite{HEIT54,MUS91} 
\begin{align}\label{Heitler1}
D=&\frac{1}{(p^--p_j^{\prime-})(p^--p_i^--k_1^-)(p^--p_k^--k_2^-)}\nonumber\\
=&\int dp^{\prime-}\delta(p^{\prime-}-p^-) \frac{\mathcal P} {(p^{\prime-}-p_j^{\prime-})(p^{\prime-}-p_i^--k_1^-)(p^{\prime-}-p_k^--k_2^-)},
\end{align}
Using the relation between distributions
\begin{align}
\frac{\mathcal P}{(p^{\prime-}-p_j^{\prime-})}\delta(p^{\prime-}-p^-)=-{1\over2} \delta^\prime(p^{\prime-}-p^-).
\end{align}
and integrating by parts we obtain
\begin{align}\label{Heitler2}
D=&{1\over2}\int dp^{\prime-}\delta(p^{\prime-}-p^-){d\over dp^{\prime-}}\biggl[{1\over{(p^{\prime-}-p_i^--k_1^-)(p^{\prime-}-p_k^--k_2^-)}}\biggr]
\nonumber\\
=&-\frac{1}{2(p^--p_i^--k_1^-)^2(p^--p_k^--k_2^-)}-\frac{1}{2(p^--p_i^--k_1^-)(p^--p_k^--k_2^-)^2}
\end{align}
Thus, in limit I, Eq.~(\ref{delta3c}) becomes
\begin{align}
(\delta m^2)_{4c}^{PV}=&{e^4\over{(2 \pi)^6}} \int {{d^2{\bf k}_{1\perp}}{d^2{\bf k}_{2\perp}} }\int {{{dk_1^+}\over{k_1^+}}{{dk_2^+}\over {k_2^+}}}\frac{2(p\cdot \epsilon(k_1))^2(p\cdot \epsilon(k_2))^2-(p\cdot k_2)(p\cdot \epsilon(k_1))^2}{8p^+}\nonumber\\&\biggl[\frac{p^+}{(p\cdot k_1)^2 (p\cdot k_2)} +\frac{p_k^+}{(p\cdot k_1)(p\cdot k_2)^2}\biggr]
\end{align}
The traces are calculated using Mathematica. 
\section{Transition matrix element for self energy in coherent state basis}
The contribution corresponding to diagrams in Fig. 6 is given by 
\begin{displaymath}
(\delta m^2)_4^\prime=(\delta m^2)_{6a}^\prime+(\delta m^2)_{6b}^\prime+(\delta m^2)_{6c}^\prime
\end{displaymath}

\begin{align}\label{delta4a}
{(\delta m^2)}_{6a}^\prime=&\frac{e^4}{2(2\pi)^6}\int{{d^2{\bf k}_{1\perp}}{d^2{\bf k}_{2\perp}}}\int{{dk_1^+}{dk_2^+} \over {16k_1^+k_2^+}}\nonumber\\ &\frac{Tr[\not\epsilon^{\lambda_2}(k_2)(\not p_k+m)\not\epsilon^{\lambda_1}(k_1)(\not p_i+m)\not\epsilon^{\lambda_1}(k_1)(\not p+m)] (p\cdot\epsilon^{\lambda_1}(k_1))\Theta_\Delta(k_1)}{(p\cdot k_1)^2[(p\cdot k_1)+(p\cdot k_2)-(k_1\cdot k_2)]}
\end{align}
\begin{align}\label{delta4b}
{(\delta m^2)}_{6b}^\prime=&\frac{e^4}{2(2\pi)^6}\int{{d^2{\bf k}_{1\perp}}{d^2{\bf k}_{2\perp}}}\int{{dk_1^+}{dk_2^+} \over {16k_1^+k_2^+}}\nonumber\\&\frac{Tr[\not\epsilon^{\lambda_2}(k_2)(\not p_k+m)\not\epsilon^{\lambda_1}(k_1)(\not p_k+m)\not\epsilon^{\lambda_2}(k_2)(\not p+m)](p\cdot\epsilon^{\lambda_1}(k_1))\Theta_\Delta (k_1)}{(p\cdot k_1)(p\cdot k_2)[(p\cdot k_1)+(p\cdot k_2)-(k_1\cdot k_2)]}
\end{align}
Calculating the trace, adding Eqs.~(\ref{delta4a}) and (\ref{delta4b}) and taking limit I, we obtain  Eq.~(\ref{coh4}).

\begin{align}\label{delta4c}
{(\delta m^2)}_{6c}^\prime=-&\frac{e^4}{2(2\pi)^6}\int{{d^2 {\bf k}_{1\perp}}{d^2 {\bf k}_{2\perp}}}\int {{dk_1^+}{dk_2^+}\over {32k_1^+k_2^+p_i^+p_j^+p_k^+}}\nonumber\\ &\frac{Tr[\not\epsilon^{\lambda_2}(k_2)(\not p_k+m) \not\epsilon^{\lambda_2}(k_2) (\not p_j^\prime+m)\not\epsilon^{\lambda_1}(k_1)(\not p_i+m)\not\epsilon^{\lambda_1}(k_1)(\not p+m)]} {(p^--p_i^--k_1^-)(p^--p_j^{\prime -})(p^--p_k^--k_2^-)}
\end{align}
Now using Heitler method \cite{HEIT54}, we obtain Eq.~(\ref{coh4c}).

\end{document}